\documentclass[pre,superscriptaddress,twocolumn,showkeys,showpacs]{revtex4-1}
\usepackage{graphicx}
\usepackage{latexsym}
\usepackage{amsmath}
\begin {document}
\title {Coexistence  and critical behaviour in a lattice model of competing species}
\author{Jacek Wendykier}
\author{Adam Lipowski}
\affiliation{Faculty of Physics, Adam Mickiewicz University, 61-614
Pozna\'{n}, Poland}
\author{Ant\'onio  Luis Ferreira}
\affiliation{Departamento de  Fisica and I3N, Universidade de Aveiro, 3810-193 Aveiro, Portugal}
\begin {abstract}
In the present paper we study a lattice model of two species competing for the same resources.  Monte Carlo simulations for $d=$1, 2, and 3 show that when resources are easily available both species coexist. However, when the supply of resources is on an intermediate level, the species with slower metabolism becomes extinct. On the other hand, when resources are scarce it is the species with faster metabolism that becomes extinct. The range of coexistence of the two species increases with dimension.  We suggest that our model might describe some aspects of the competition between normal and tumor cells. With such an interpretation, examples of tumor remission, recurrence and of different morphologies are presented. In the $d=1$ and $d=2$ models, we analyse the nature of phase transitions: they are either discontinuous or belong to the directed-percolation universality class, and in some cases they have an active subcritical phase. In the $d=2$ case, one of the transitions seems to be characterized by critical exponents different than directed-percolation ones, but this transition could be also weakly discontinuous. In the $d=3$ version, Monte Carlo simulations are in a good agreement with the solution of the mean-field approximation. This approximation predicts that oscillatory behaviour occurs in the present model, but only for $d\gtrsim 2$. For $d\geq 2$, a steady state depends on the initial configuration in some cases.
\end{abstract}

\maketitle
\section{Introduction}
Competition is a fundamental force shaping almost every biosystem \cite{begon}. It operates at the level of species, and leads to development of new adaptations and creation or extinction of  species. The competition occurs also among individuals of the same species, e.g., when two neighbouring plants have to share the same resources like water  or nutrients in the soil. Also at the cellular level the competition is present, but various mechanisms usually keep it under control. These mechanisms sometimes, however,  fail and a group of tumor cells with abnormal reproduction and differentiation pattern emerges. The growth of tumor is a very complex process since tumor cells compete with normal cells for food, space or waste removal, modify vascular system or other tissues, and often lead to the death of an organism~\cite{crespi,merlo}. Therefore, realistic models of tumors should take into account numerous factors. Due to this complexity, the multiscale modeling is very often used \cite{alaracon}, but even relatively simple models that use generalized Lotka-Volterra equations \cite{sarkar} are difficult to analyse and understand.

Since tumor development can be regarded as a spatio-temporal pattern-formation process, cellular automata seem to provide a natural platform to model such phenomenon \cite{moreira,khainepl,simpson,simpson2007,khainjsp,bearer,oden}.  This approach was used, for example, to study general aspects of Gompertzian tumor-growth problem \cite{qi}, or more realistic  analysis of the three-dimensional brain-tumor model on Voronoi tessallation \cite{kansal}. Some hybrid models that combine cellular automata with partial differential equations were also used to describe interactions
between a tumor and the immune system of the host organism \cite{mallet}, or morphologies of an avascular tumor \cite{ferr2002}.  One should also mention that related mathematical and computational approaches have been used to model of some properties of tissues or  genomes \cite{paulsonn, gabetta, bellomo, bellomo2010}.

The cellular automata implemented in such models are also very complex, often with non-local, heterogeneous or state-dependent transition rules. It is thus very difficult to understand their behaviour on general grounds. And some basic insight into the behaviour of such systems would be indeed desirable since it could help us to understand the role of for example fluctuations and competition in such systems.  Fluctuations and competition play major role in determining the symmetry and the nature of the ordering in many statistical mechanics systems. They are also known to have spectacular consequences in chemical or biological systems \cite{lefever}. To explain phenomena like spontaneous cancer remission, that puzzles medicine for decades \cite{everson},  the coupling of fluctuations and competition certainly has to be understood.

Cellular automata can also be successfully applied in ecology, where competition and fluctuations are of great  importance. One of the classical problems in this field is related to the competitive exclusion principle formulated fifty years ago by Gause \cite {gause}. It states that when different species compete for the same resources, only the fittest one survives. However, this principle seems to be in contradiction with the observed abundance of species coexisting in ecosystems, of which plankton is a particularly evident example \cite{hutchinson}. Some resolutions of this so-called plankton paradox were proposed referring to spatio-temporal inhomogeneities \cite{miyazaki}, cyclic dominance \cite{reichenbach}, or size-selective grazing \cite{wiggert}. Taking into account the basic nature and potential importance of Gause's principle and a variety of interactions in ecosystems, it would be desirable to examine this problem also in some other models.

In the present paper we examine a simple lattice  model of two species competing for renewable resources. The model can be regarded as a generalization of a certain lattice prey-predator model \cite{lip1999}.  Such models can be classified as some kind of cellular automata, but due to their asynchronous dynamics they are also closely related to some other statistical-mechanics models, namely particle systems \cite{odor}. The species in our model obey the intraspecific (but not interspecific) exclusion rule. Using Monte Carlo simulations and a mean-field approximation, we examine  phase diagrams and phase transitions, which mark the limits of existence or coexistence of species. The Monte Carlo simulations are made for one-, two-, and three-dimensional versions of the model, while the mean-field approximation enabled us to make some predictions concerning the coexistence of species or the oscillatory behaviour even for higher-dimensional models.  The two species in our model can also be interpreted as normal and tumor cells and actually, we will use such a terminology throughout our paper. In this context we show examples of tumor remission and recurrence, and of different morphologies of invading tumors. One has to be aware, however, that such an interpretation must be taken with considerable care since the present model provides only a very crude description of competition between normal and tumor cells.

\section{Model}
Lattice models are frequently used to describe various problems in ecology \cite{grimm}.
Such individual-based approach very often supplements more traditional techniques based on differential equations \cite{murray}. A classical problem in this field, having origin in the works of Lotka and Volterra, are  prey-predator systems \cite{pekalski,tome,albano}. An intensive research, especially in the physicists community,  is inspired  by the fact that the dynamics of lattice models of such systems generates fluctuations that can wash-out spatio-temporal patterns predicted by coarse-grained approaches \cite{durrett,mobilia2007,aguiar}.

In our model two species, which differ in the metabolic and reproduction rates, occupy lattice sites and consume the same renewable resources.
In the context of tumor-growth problem, faster evolving species can be considered as tumor or cancer (c), the slower one as normal cells (n), and resources (p) become nutrients provided with blood.
Each site of a \mbox{$d$-dimensional} lattice of linear size $L$ can be empty or occupied by a nutrient, by a normal cell, or by a tumor cell (and we have implemented periodic boundary conditions). Moreover, we implement the intraspecific exclusion rule: no more than one cell of a given type can occupy a given site. It means that in our model each site can be in one of the eight states $(p,n,c)$, where $p,n,c=0$ or 1. For example, (0, 1, 1) stands for the absence of a nutrient cell and the presence of both normal and tumor cells.
The competition between normal and tumor cells for nutrients is modeled in a way that resembles prey-predator systems with n and c being two predator species, and p playing the role of preys. As a matter of fact, our model can be regarded as an extension of the prey-predator model that was already examined in the context of oscillatory and critical behaviour that such systems are known to exhibit \cite{lip1999,KOWALIK}.

The detailed rules of the model are specified below:
\begin{itemize}
\item Choose a site randomly. 
\item With probability $r_p$  update the site provided that there is a nutrient cell on this site ($p=1$). In this case an update means an attempt to breed: one of the nearest neighbours of the selected site, not containing a nutrient cell, is chosen and a new nutrient cell is placed there.

\item With probability $r_n$ update the site provided that there is a normal cell there ($n=1$). In this case the normal cell survives only if there is also a nutrient cell on this site. If so, the normal cell consumes the nutrient cell ($(p=1)\rightarrow (p=0)$) and it attempts to breed (in a way analogous to that described above for a nutrient cell).
\item With probability $r_c$ update the site provided that there is a tumor cell there ($c=1$). The tumor cell survives only if there is also a nutrient cell on this site. If so, the tumor cell consumes the nutrient cell and attempts to breed in an analogous way as a nutrient cell does.
\end{itemize}

The update probabilities are the only control parameters of the model and they satisfy the obvious condition: 
\begin{equation}
r_p + r_n + r_c =  1. 
\label{normalization}
\end{equation}

We assume that tumor cells have faster metabolism and reproduction rates than normal cells (i.e., $r_c>r_n$). With Eq.~(\ref{normalization}) and a fixed ratio of $r_c/r_n$, the behaviour of the model depends only on~$r_p$. 

To examine the model, we introduce eight densities $x_{pnc}$, which measure the average concentrations of the respective states in a system:
\begin{equation}
x_{pnc}=\frac{1}{L^d}\sum_i \delta(i,(p,n,c)),
\label{densities}
\end{equation}
where summation is over all lattice sites $i$, and $\delta(i,(p,n,c))=1$ when the site $i$ is in the state $(p,n,c)$, and 0 otherwise.
Using the densities, we can calculate the concentrations of nutrient ($x_p$), normal ($x_n$), and tumor cells ($x_c$) as follows:
\begin{eqnarray}
x_p & = & x_{100}+x_{110}+x_{101}+x_{111}, \nonumber\\
x_n & = & x_{010}+x_{110}+x_{011}+x_{111}, \nonumber\\
x_c & = & x_{001}+x_{101}+x_{011}+x_{111}.\label{xpnc}
\end{eqnarray}
The densities  $x_p$, $x_n$, and $x_c$ are quantities of our main interest.

Although different metabolism and reproduction rates inclined us to interpret two species in our model as normal and tumor cells, such an analogy must be taken with considerable care. Indeed, a number of features of real normal-tumor  competition is not properly reflected in our model. For example, normal and tumor cells cannot occupy the same place, proliferation of normal cells is extremely small comparing to the cancer cells, and the latter ones in some cases might even diffuse. Moreover, nutrients are much smaller than normal or tumor cells and they do not divide.
We are aware of such shortcomings of our model, but the objective of understanding of competing systems from the perspective of statisitcal mechanics implied that a number of factors had to be omitted. As we will show in Section V some of the properties of our model seem to be relatively robust, and there is a hope that more realistic systems will exhibit similar behaviour.

\section{Mean-field approximation}
For the model defined by the above dynamical rules, one can write a Master equation and solve it using some decoupling procedure \cite{tome,ferreira}. The first step in this approach is to sum over states of all but one site. As a result we obtain equations describing the time evolution of probabilities of a single-site configuration. Such equations will contain probabilities of more complex configurations (i.e. two-site configurations) that need to be subsequently factorized (i.e., approximated with some products of probabilities of single-site configurations). Eventually, each term contributing to the evolution of single site probabilities will contain probability that a chosen site is in an appropriate state (for  a given process) and the rate of that process ($r_p,\ r_n$ or $r_c$). In some processes neighbouring sites also give some contributions. For our model one arrives at the following equations, describing the time evolution of the densities~$x_{pnc}$:
{\allowdisplaybreaks
\begin{eqnarray}
\frac{dx_{100}}{dt} & = & r_px_pf(x_p)x_{000}-r_n\tilde{x_n}f(x_n)x_{100}-\nonumber\\
&  & r_c\tilde{x_c}f(x_c)x_{100},\nonumber \\
 \frac{dx_{010}}{dt} & = & r_n\tilde{x_n}f(x_n)x_{000}+r_cx_{011}+r_n(x_{110}-x_{010})-\nonumber\\
&  & r_px_pf(x_p)x_{010}-r_c\tilde{x_c}f(x_c)x_{010},\nonumber\\
 \frac{dx_{001}}{dt} & = & r_c\tilde{x_c}f(x_c)x_{000}+r_nx_{011}+r_c(x_{101}-x_{001})-\nonumber\\
&  & r_px_pf(x_p)x_{001}-r_n\tilde{x_n}f(x_n)x_{001},\nonumber\\
 \frac{dx_{110}}{dt} & = & r_px_pf(x_p)x_{010}+r_n\tilde{x_n}f(x_n)x_{100}-\nonumber\\
&  & x_{110}[r_c\tilde{x_c}f(x_c)+r_n],\nonumber\\
 \frac{dx_{101}}{dt} & = & r_px_pf(x_p)x_{001}+r_c\tilde{x_c}f(x_c)x_{100}-\nonumber\\
&  & x_{101}[r_n\tilde{x_n}f(x_n)+r_c],\nonumber\\
 \frac{dx_{011}}{dt} & = & (r_n+r_c)x_{111}+r_n\tilde{x_n}f(x_n)x_{001}+\nonumber\\
&  & r_c\tilde{x_c}f(x_c)x_{010}-x_{011}[r_px_pf(x_p)+r_n+r_c],\nonumber\\
 \frac{dx_{111}}{dt} & = & r_n\tilde{x_n}f(x_n)x_{101}+r_c\tilde{x_c}f(x_c)x_{110}+\nonumber\\
&  & r_px_pf(x_p)x_{011}-x_{111}(r_n+r_c),\label{mfa}
\end{eqnarray}
}
where the unit of time corresponds to a single (on average) update per lattice site and
\begin{eqnarray}
\tilde{x_n} & = & x_{110}+x_{111},\nonumber\\
\tilde{x_c} & = & x_{101}+x_{111}.
\end{eqnarray}
Moreover, the function $f(x)$ denotes the average number of sites that had chosen to breed (with respect to $x$) at a given site
\begin{equation}
 f(x)=w\sum_{k=0}^{2d-1} \frac{1}{k+1}\frac{(2d-1)!}{k! (2d-1-k)! } (1-x)^k x^{2d-1-k}\label{fx}
\end{equation}
where $w$ denotes the number of nearest neighbors (on cartesian lattices $w=2d$). 
After simple algebra one can show that $f(x)=\frac{1-x^w}{1-x}$.
The quantities $\tilde{x_n}$ ($\tilde{x_c}$) denote the densities of sites with both normal (tumor) and nutrient cells.  The derivative  $ \frac{dx_{000}}{dt}$ obeys a similar equation, but it is simpler to use the obvious normalization condition
\begin{equation}
\sum_{p,n,c=0,1} x_{pnc}=1.
\label{normalizx}
\end{equation}

That mean-field equations (\ref{mfa}) are indeed consequences of dynamical rules can be seen also from less formal arguments. For example, the first term in the first equation describes the increase of $x_{100}$ due to the breeding of nutrient cells whose neighbour chosen for the placement of a newborn nutrient cell is empty. The factor $f(x_p)x_{000}=(1-x_p^w)\cdot \frac{x_{000}}{1-x_p}$ gives the probability that at least one of the neighbouring sites does not contain a nutrient cell ($(1-x_p^w)$) and that the chosen neighbouring  site (that does not contain a nutrient cell) is in the state $(0,0,0)$ (the conditional probability of this equals~~$\frac{x_{000}}{1-x_p}$).

We solved numerically the equations (\ref{mfa}) and calculated the densities (\ref{xpnc}). The results of our calculations together with those of Monte Carlo simulations are discussed below. Let us also notice that for some biologically-inspired lattice models (i.e., wound healing problem) other approximate descriptions are possible based for example on Cahn-Hilliard equation \cite{khain2008}.

\section{Results of Monte Carlo simulations}

Monte Carlo simulations are often used to study lattice models. Starting for each $r_p$ from a new random configuration (unless specified otherwise, we used configurations where each site with probability 1/4 was either empty or occupied by a nutrient, normal, or tumor cell), we measured the steady-state densities (\ref{xpnc}) as functions of $r_p$, for the fixed ratio $r_c/r_n$.  Of course, before measurements we relaxed the system and monitored some densities, until it reached the steady state. To examine critical behaviour and the nature of some phase transitions in the model, we also measured the time dependence of some of these densities. Additional details concerning our simulations are provided in the following subsections.

\subsection{d=1}

\begin{figure}
\includegraphics[width=9cm]{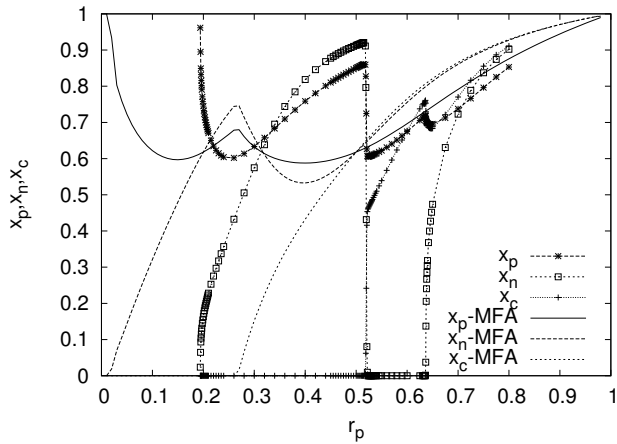}  \vspace{0cm} \caption{ \label{steadyd1}Steady-state densities of nutrient ($x_p$), normal ($x_n$), and cancer ($x_c$) cells as functions of~~$r_p$ for $d=1$ and $r_c/r_n=3$. The simulations were made for $L=10^5$ and simulation time $t=10^4$. Close to critical points, longer simulations were performed with $t=10^6$. For each value of $r_p$, an initial configuration was random and contained individuals of all species. Solutions of the mean-field approximation (\ref{mfa}) are also shown, but, as one can see, the agreement with simulations is rather poor.}
\end{figure}

The plot of the steady-state densities $x_p,\ x_n,$ and $x_c$ as functions of $r_p$ for $r_c/r_n=3$ is shown in Fig.~\ref{steadyd1}.  For large $r_p$ mainly nutrient cells are updated, i.e., they breed if there is a place nearby. As a result, nutrients are easily available and both normal and tumor cells can coexist. For smaller values of $r_p$, the shortage of nutrients affects mainly normal cells and for $r_p\sim 0.6367$ they become extinct. For $r_p$ even smaller, only tumor cells can exist -- although an initial configuration always contains individuals of both species, normal cells quickly die out, and only  tumor cells survive. Then, surprisingly, a drastic change takes place: for $r_p\lesssim0.5195$ these are normal cells that survive and tumor cells that become extinct. When $r_p$ is too small (for $r_p\lesssim0.1942$), nutrients do not reproduce sufficiently fast and both normal and tumor cells become extinct (and only nutrients survive).

As might be expected, for low-dimensional systems the solution predicted by the mean-field approximation (\ref{mfa}) with $w=2$ considerably differs from Monte Carlo simulations (see Fig.~\ref{steadyd1}). 
\begin{figure}
\includegraphics[width=\columnwidth]{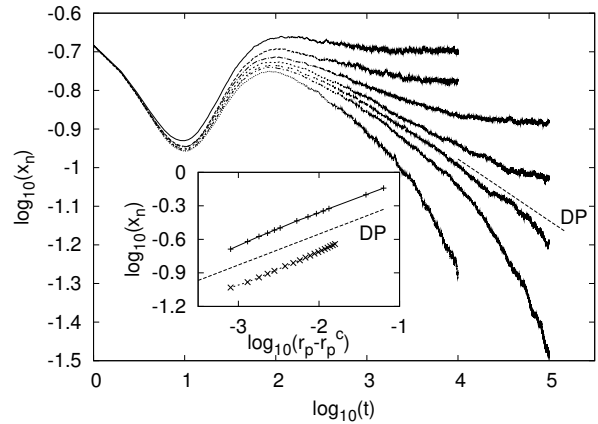}  \vspace{0cm} \caption{The time dependence of the density of normal cells $x_n$ calculated for $d=1$, $r_c/r_n=3$, and (from top) $r_p=$ 0.205, 0.2, 0.197, 0.195, 0.194, 0.193, and 0.19. The dotted line has a slope corresponding to the (1+1) directed percolation value, $\delta_{DP}=0.1595$ \cite{hinrichsen}. The inset shows the behaviour of $x_n$ in the vicinity of critical points located at $r_p=0.1942$ (+) and $r_p=0.6367$ ($\times$). In both cases, power-law behaviour is in a good agreement with the directed percolation value, $\beta_{DP}=0.2765$.} \label{decay1942}
\end{figure}

The limits of existence of normal or tumor cells are marked by phase transitions in the model. To examine their nature, we calculated the time-dependence of the average density of species that becomes extinct at the transition point (usually the averages are taken over $10^2$ independent runs, which start from different random initial configurations).
One expects that at the critical point these quantities show power-law decay $x(t)\sim t^{-\delta}$, where the characteristic exponent $\delta$ exhibits some degree of universality \cite{hinrichsen}. The results of our calculations for the transition at $r_p=0.1942$ are shown in Fig.~\ref{decay1942}, and from the behaviour at the critical point we estimate $\delta\sim 0.16$.  This value is very close to the corresponding exponent in the directed percolation (DP) problem, $\delta_{DP}=0.1595$ \cite{hinrichsen}. An additional argument that the model belongs to the DP universality class comes from the analysis of the steady-state density of normal cells $x_n$ close to the critical point. We find that $x_n\sim (r_p-0.1942)^{\beta}$, where $\beta$ is very close to the directed percolation value, $\beta_{DP}=0.2765$ \cite{hinrichsen} (see the inset in Fig.~\ref{decay1942}).
The transition at $r_p=0.1942$ is similar to some other transitions in models with absorbing states \cite{odor,hinrichsen}. The critical behaviour of the DP universality class is typical for a large class of models  with a single absorbing state and our model also falls into this class \cite{comment1}.

Similar calculations were performed for the transition at $r_p=0.6367$, where normal cells become extinct. We show only the steady-state values of the density of normal cells  (inset in Fig.~\ref{decay1942}) and they suggest that also in this case the model most likely belongs to the DP universality class. 
The time dependence of $x_n$ at the critical point (not presented) yields the estimation of $\delta$,  which is also consistent with the DP value.
Let us notice, however, an important difference between the phase transition at 0.1942, where the model enters an absorbing state, and the one at 0.6367, where instead of entering the absorbing state, the system remains in the active phase with tumor and nutrient cells (but without normal cells). This phase transition provides yet another example that the DP critical behaviour appears for a wider class of models than those with a single absorbing state. However, a complete understanding of this issue is not yet achieved \cite{munoz,wijland,park}.
\begin{figure}
\includegraphics[width=\columnwidth]{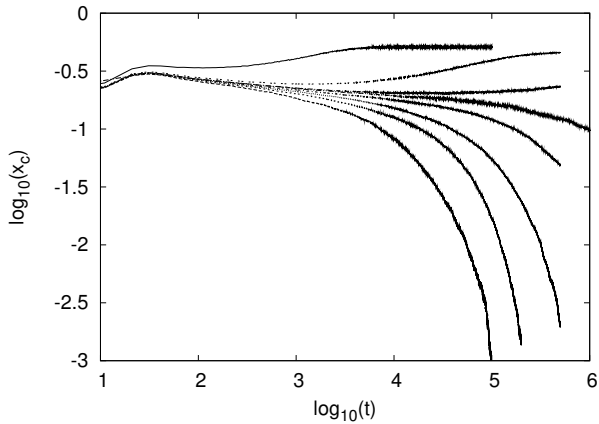}  \vspace{0cm} \caption{The time dependence of the density of tumor cells $x_c$ calculated for $d=1$, $r_c/r_n=3$, and (from top) $r_p=$ 0.54, 0.522, 0.52, 0.5195, 0.519, 0.518, 0.517, and 0.515.} \label{decay052}
\end{figure}

A different behaviour is observed for the transition at $r_p=0.5195$. The steady-state densities exhibit pronounced jumps (Fig.~\ref{steadyd1}) and that suggests a discontinuous transition at this point. The plot of time dependence of the tumor cells confirms such behaviour (Fig.~\ref{decay052}). Indeed, no power-law  decay of $x_c$ as a function of time is seen at the critical point.

The above-described behaviour seems to be generic in one-dimensional version of our model. We observed  the same sequence of phase transitions (DP-discontinuous-DP), albeit at other values of $r_p$,  for $r_c/r_n$ ranging from 1.1 to 10. It would be interesting to find a reason of such a robust behaviour and we will return to this problem in the next subsection.
\begin{figure}
\vspace{1.5cm}
\includegraphics[width=\columnwidth]{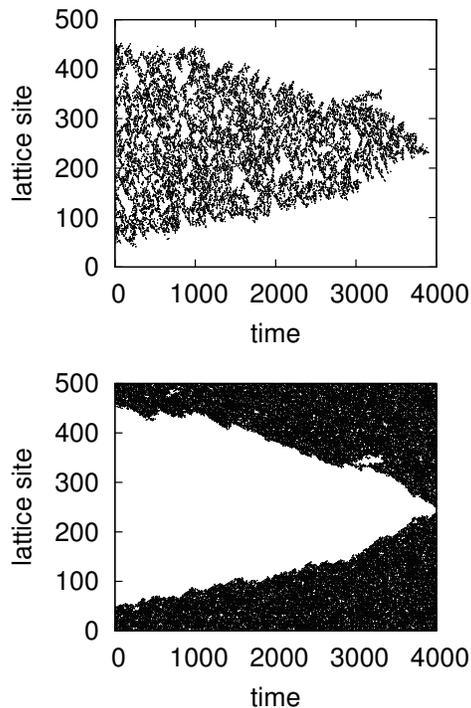}  \caption{Tumor remission. The time evolution of distribution of tumor (top) and normal (bottom) cells calculated for $d=1$, $r_c/r_n=3$, and $r_p=0.48$. Sites occupied by tumor (top) or normal (bottom) cells are plotted with black dots.} 
\label{conf-time}
\end{figure}

\begin{figure}
\vspace{1.5cm}
\includegraphics[width=\columnwidth]{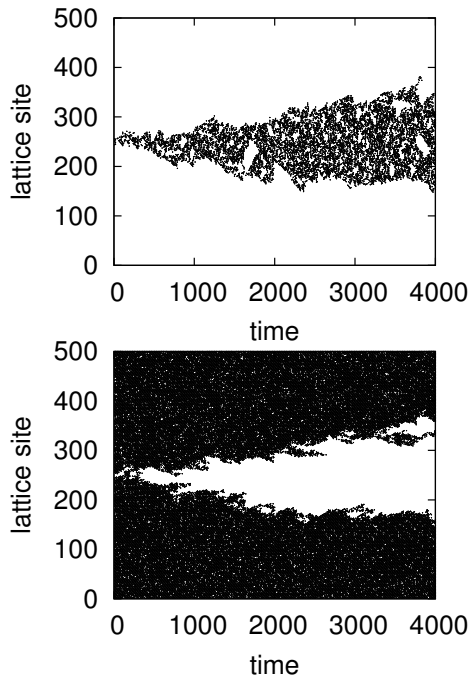}  \caption{Tumor recurrence. The time evolution of the distribution of tumor (top) and normal (bottom) cells calculated for $d=1$, $r_c/r_n=3$, and $r_p=0.55$. Sites occupied by tumor (top) or normal (bottom) cells are plotted with black dots.} 
\label{conf-timea}
\end{figure}

In studies on tumor growth, it is often important to examine the time evolution of frequency or distribution of tumor cells \cite{merlo}. The observed patterns of invasion of tumor cells sometimes resemble a scenario known to ecologists as "seed and soil", where a new species colonizes a new habitat \cite{fidler}. 
In our model a qualitatively similar analysis can be made. In Fig.~\ref{conf-time} we show an example of tumor remission. In the initial configuration,  80\% of the central sites were occupied by nutrient and tumor cells (1,0,1) and the rest by nutrient and normal cells (1,1,0). The simulations were performed for $r_c/r_n=3$ and $r_p=0.48$, for which we have only nutrient and normal cells in the steady state. In Fig.~\ref{conf-time} a clear extinction of tumor cells takes place. An opposite effect is seen in Fig.~\ref{conf-timea} where the initial configuration contained only a small cluster of 10 tumor cells, while $r_p=0.55$. Under such conditions a tumor recurrence can be observed with normal cells becoming gradually extinct. 

\subsection{d=2}

\begin{figure}
\includegraphics[width=9cm]{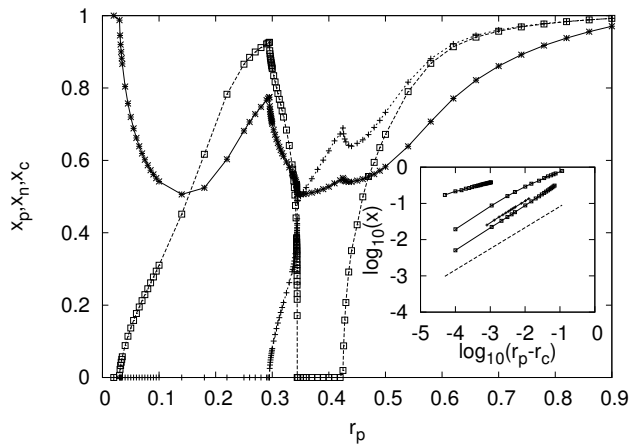}  \vspace{0cm} \caption{ \label{steadyd2}Steady-state densities of nutrient ($x_p$), normal ($x_n$), and cancer ($x_c$) cells as functions of $r_p$ for $d=2$ and $r_c/r_n=3$. Simulations were made for $L=500$ and simulation time $t=10^4$. Close to critical points, longer simulations were performed with $t=10^6$. The inset shows the behaviour of $x_n$ and $x_c$ in the vicinity of critical points located at (from top) $r_p=0.3441$, 0.4249, 0.2947, and 0.0299. Except for $r_p=0.3441$, the power-law behaviour is in a good agreement with the (2+1) directed percolation value, $\beta_{DP}=0.584$  (dashed line) \cite{hinrichsen}.}
\end{figure}

We made similar calculations for the two-dimensional model. For $r_c/r_n=3$, steady-state densities are shown in Fig.~\ref{steadyd2}. Although there are similarities to the $d=1$ case, there are also some qualitative differences. In particular, a single discontinuous transition that reverses the extinction of  tumor and normal cells is replaced in the $d=2$ version by two transitions and a narrow range of a coexistence: $0.2947<r_p<0.3441$ (see Fig.~\ref{steadyd2}). A more detailed analysis of the densities close to the phase transitions shows that three of these phase transitions are likely to belong to the DP(2+1) universality class (see inset in Fig.~\ref{steadyd2}). However, the scaling of the density of normal cells close to the phase transition at $r_p=0.3441$ seems to scale with the exponent $\beta \sim 0.32$, which is much smaller than the DP(2+1) exponent $\beta_{DP}=0.584$~\cite{hinrichsen}.

\begin{figure}
\includegraphics[width=\columnwidth]{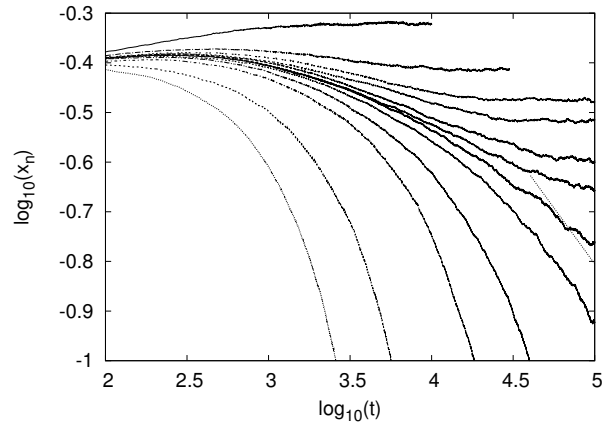}  \vspace{0cm} \caption{The time dependence of the density of normal cells $x_n$ calculated for $d=2$, $r_c/r_n=3$, and (from top) $r_p=$ 0.341, 0.343, 0.3435, 0.3437, 0.3439, 0.344. 0.3441 (thick line), 0.3442, 0.3445, 0.345, 0.347, and 0.35. The dotted line has a slope corresponding to the (2+1) directed percolation value, $\delta_{DP}=0.451$ \cite{hinrichsen}.} \label{decay344}
\end{figure}

An additional argument that this phase transition might be described by exponents different than DP(2+1)  is presented in Fig.~\ref{decay344}, which shows on the log-log scale the time dependence of the density of normal cells $x_n$. At the transition point (thick line), the asymptotic decay of $x_n$ seems to be characterized by the exponent $\delta\approx 0.25$, which is different from $\delta_{DP}=0.451$~\cite{hinrichsen}. We do not present numerical data here, however, for the remaining three transitions the estimated exponent $\delta$ was very close to the DP(2+1) value.

Despite such a discrepancy with the expected DP(2+1) universality class, our estimations of critical exponents must be taken with some care. Namely, it is known that such estimations  for nonequilibrium phase transitions are often  very difficult and we cannot exclude that much more extensive numerical calculations will modify our estimations. Let us also notice that the phase transition at $r_p=0.3441$ has an active subcritical phase (nutrient and tumor cells), and such a feature might perhaps cause numerical difficulties or even change the critical behaviour (however, the phase transition at $r_p=0.4249$ seems to have the DP(2+1) critical exponents, but also have an active subcritical phase).

\begin{figure}
\includegraphics[width=\columnwidth]{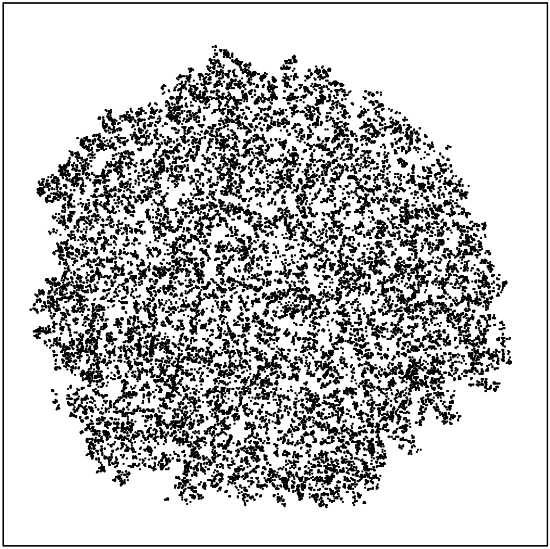}  \vspace{0cm} \caption{The snapshot distribution of tumor cells calculated for $d=2$, $r_c/r_n=3$, $r_p=0.304$ and simulation time $t=10^4$. The initial configuration contained a small seed of tumor cells surrounded by normal and nutrient cells.} \label{conf304a}
\end{figure}

\begin{figure}
\includegraphics[width=\columnwidth]{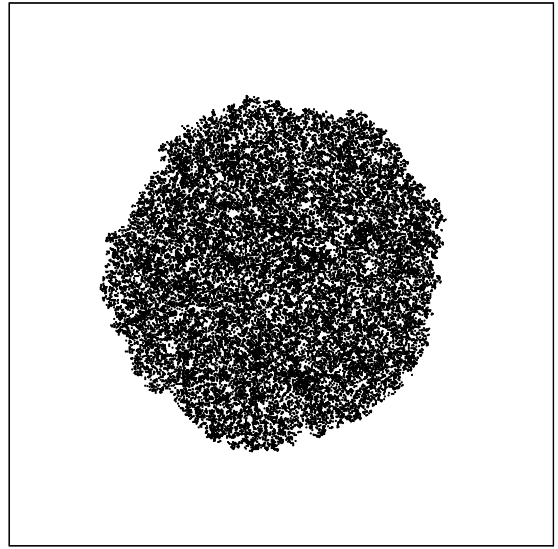}  \vspace{0cm} \caption{The snapshot distribution of tumor cells calculated for $d=2$, $r_c/r_n=3$, $r_p=0.34$,  and simulation time $t=2\cdot 10^3$. The initial configuration contained a small seed of tumor cells surrounded by normal and nutrient cells.} \label{conf34a}
\end{figure}

Tumors in two-dimensional models have much richer morphology than those in the one-dimensional case. Figs.~\ref{conf304a}--\ref{conf34a} show the distribution of tumor cells that grew from a small tumor seed surrounded by normal and nutrient cells. Let us notice that for $r_p=0.34$, which is relatively far from the limit of existence of tumor (0.2947), the tumor remains compact and dense (Fig.~\ref{conf34a}). For $r_p=0.304$, i.e., very close to such a limit of existence, the shape of tumor is much more irregular (Fig.~\ref{conf304a}). Morphology of tumors is very important from the clinical point
 of view, however, a precise comparison with existing data is very often difficult~\cite{schaller}. Nevertheless, certain morphologies can be associated with some forms of cancer \cite{ferr2002}. 

\begin{figure}
\includegraphics[width=\columnwidth]{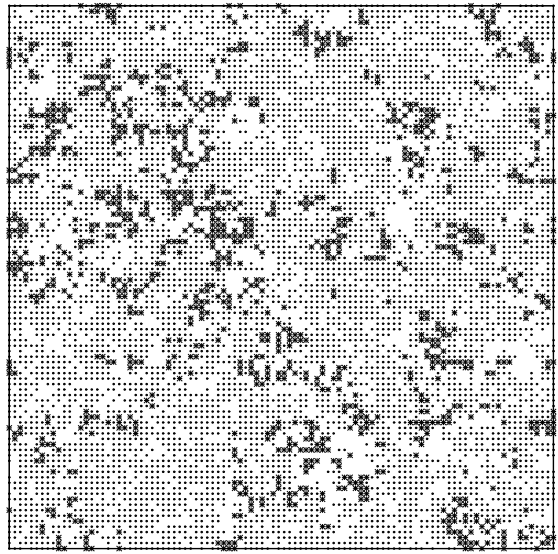}  \vspace{0cm} \caption{The snapshot distribution of nutrient (\textperiodcentered) and tumor (\textasteriskcentered) cells calculated for $d=2$, $r_c/r_n=3$, and $r_p=0.305$. Tumor cells seem to concentrate close to empty regions -- apparently in the interior they lose the competition with normal cells.} \label{conf305}
\end{figure}

Let us notice that the coexistence of tumor and normal cells in the range $0.2947<r_p<0.3441$ is very fragile since this interval is rather narrow and both species remain close to their limits of existence. To have some insight into the nature of this coexistence, we present the distribution of  nutrient and tumor cells (Fig.~\ref{conf305}). We can observe that tumor cells seem to concentrate close to  empty regions. Apparently,  under such conditions, in the interior they lose the  competition with normal cells. It is tempting to speculate that such structures might have a rather short lifetime in one-dimensional systems and that is why there is no coexistence for the $d=1$ model except in the large $r_p$ regime (see Fig.~\ref{steadyd1}).

It turns out, however, that the behaviour in the $d=2$ version, which we have shown in Fig.~\ref{steadyd2}, is not entirely generic and to some extent depends on the ratio $r_c/r_n$. Namely, numerical simulations (which we do not present) show that for $r_c/r_n=1.5$ and 1.1 some of the continuous transitions become discontinuous, as in one-dimensional system, but with a narrow range of $r_p$ for which both tumor and normal cells coexist. For larger $r_c/r_n$ (we made simulations for $r_c/r_n=5$) all phase transitions remain continuous. Taking into account appearance of discontinuous transitions one cannot exclude that the phase transition for $r_c/r_n=3$ at $r_p=0.3441$ is actually weakly discontinuous and that is why our estimations of critical exponents do not agree with the DP universality class exponents.

\subsection{d=3}
\begin{figure}
\includegraphics[width=\columnwidth]{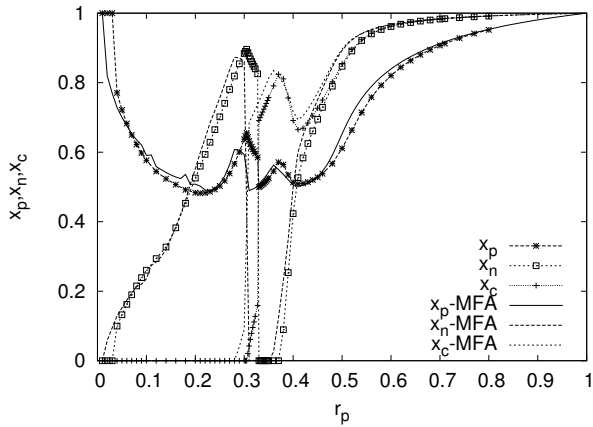}  \vspace{0cm} \caption{ \label{steadyd3}Steady-state densities of nutrient ($x_p$), normal ($x_n$), and cancer ($x_c$) cells as functions of $r_p$ for $d=3$ and $r_c/r_n=1.5$. Simulations were made for $L=50$ and simulation time $t=10^4$. Close to critical points, longer simulations were performed with $L=80$ and $t=10^5$. The solutions of the mean-field approximation (\ref{mfa}) are also shown. For $0.05 \lesssim r_p \gtrsim 0.22$ normal cells and nutrients show oscillatory behaviour and in such a case the plotted values should be interpreted as time averages rather than steady-state values.}
\end{figure}

\begin{figure}
\includegraphics[width=\columnwidth]{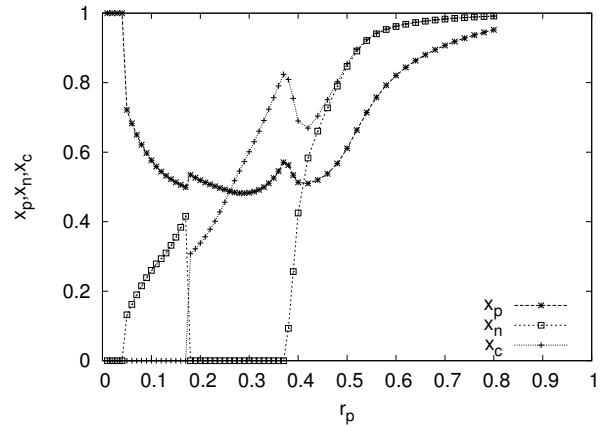}  \vspace{0cm} \caption{ \label{dific}Steady-state densities of nutrient ($x_p$), normal ($x_n$), and cancer ($x_c$) cells as  functions of $r_p$ for $d=3$ and $r_c/r_n=1.5$. Simulations were made for $L=50$ and simulation time $t=10^4$. In the initial configuration there was much more tumor than normal cells ($x_p=0.2$, $x_n=0.05$, and $x_c=0.65$). For $r_p<0.18$ or $0.33<r_p<1$, the steady-state densities are within numerical error the same as those in Fig.~\ref{steadyd3}.}
\end{figure}

We made numerical simulations also for the $d=3$ version. For $r_c/r_n=3$, the steady-state densities as functions of $r_p$ are shown in Fig.~\ref{steadyd3}. We did not estimate the critical exponents, but by analogy with low-dimensional versions, we expect that continuous transitions belong to the DP(3+1) universality class.
Let us also notice a good agreement with numerical solutions of the mean-field approximation~(\ref{mfa}) with $w=6$.
With a pronounced discontinuous transition seen in Fig.~\ref{steadyd3}, the three-dimensional model resembles the one-dimensional version. It seems to us, however, that there are some important differences between these two cases. Namely, we noticed that for $d=3$ the location of the discontinuous transition depends on the concentration of species in the initial configuration. For example, for initial configurations containing much more tumor than normal cells, the location of this transition is shifted toward much smaller $r_p$ (see Fig.~\ref{dific}). The dependence on the initial configuration appears only in some vicinity of the discontinuous transition. In Fig.~\ref{dific}, for $r_p<0.18$ or $0.33<r_p<1$ , the  steady-state densities are within numerical error the same as those in Fig.~\ref{steadyd3}. Such a dependence on the initial configuration is also reproduced in the solutions of the mean-field equations (\ref{mfa}). In our opinion it is likely that these are not only densities of species but also their spatial distribution that determines location of discontinuities. More detailed analysis of such effects would be certainly desirable.

On the other hand, numerical simulations (not presented) show that in one-dimensional systems location of discontinuity is essentially independent on the initial configuration (provided that it contains a non-zero concentration of normal and tumor cells, and nutrients). The simulations show also that in the two-dimensional version, for $r_c/r_n=1.5$ and 1.1, location of discontinuity depends on the initial configuration. For $r_c/r_n=3$ or 5 all transitions seem to be continuous and no dependence on the initial configuration was detected. It would be desirable to know whether there are some general physical arguments about the role of dimension in determining such a behaviour of our model.

\subsection{d$>$3}

Systems with long-range interactions or fast mixing due to diffusion are often well approximated by infinite-dimensional systems. It is thus of interest to examine models with $d>3$. Since Monte Carlo simulations are very demanding for such models, we performed only calculations within the mean-field approximation (\ref{mfa}). This approximation was quite satisfactory already for $d=3$ (Fig.~\ref{steadyd3}), and we expect that for $d>3$ the solutions of (\ref{mfa}) will be even more accurate. Our calculations show that for $d=4,5,\ldots, 10$ the steady-state densities behave similarly to the $d=3$ version. Location of the discontinuity was also found to depend on the initial condition.
\begin{figure}
\includegraphics[width=\columnwidth]{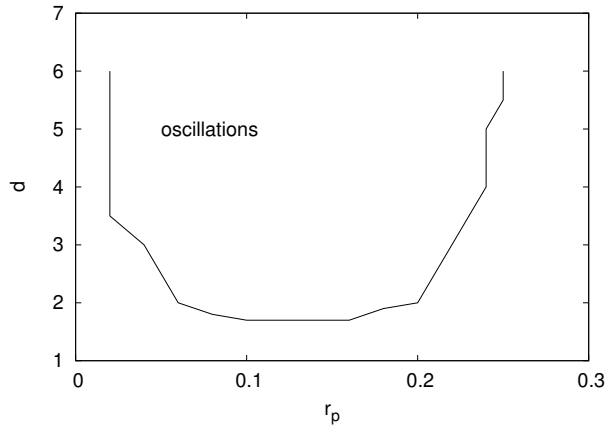}  \vspace{0cm} \caption{ \label{oscil} Boundary of oscillatory regime calculated using the mean-field approximation (\ref{mfa}) for  $r_c/r_n=3$. Oscillations occur only in the tumor-free phase of the model ($x_c=0$).}
\end{figure}

One of the interesting problems in lattice models concerns the existence of oscillations and numerous examples ranging from hetero- or homogeneous catalysis to various ecological systems have already been examined with this respect \cite{zhdanov}. On general grounds one expects that fluctuations in low-dimensional systems are strong enough to destroy such temporal oscillations in the thermodynamic limit. An important question is: what is the critical dimension $d_c$ above which such oscillations will survive in the thermodynamic limit? Grinstein~{\it et al.}~\cite{grinstein} presented certain arguments indicating that $d_c=2$. Numerical results for both synchronous \cite{chate,hemm} and asynchronous models \cite{lip1999} seem to confirm that temporal oscillations exist only for d$>$2. 

The mean-field approximation (\ref{mfa}) in some cases also predicts that solutions are oscillatory.  For a given $d$, we analysed the range of $r_p$ where the oscillatory behaviour was seen, and the corresponding plot is shown in Fig.~\ref{oscil}. Using this approximation, we estimate that $d_c\sim 1.8$ and that result is very close to the already mentioned prediction $d_c=2$.
\section {Randomly supplied nutrients}
As we already mentioned, nutrient cells in our model evolve according to rules typical rather to some living species. In particular, they divide and place their offsprings on some empty lattice sites. Such rules might be justified in modeling multi-species ecosystems where nutrients could be considered as prey-type species. However, in the context of tumor growth problems division of nutrient cells is not realistic and different rules should be used. In the present section we examine a modification of our model, where nutrient cells are supplied randomly.
More precisely, we left unchanged all the rules of the model (sect. II) except the second rule that now becomes
\begin{itemize}
\item With probability $r_p$ place a nutrient cell at the chosen site, provided that there is no such a cell there.
\end{itemize}

\begin{figure}
\includegraphics[width=9cm]{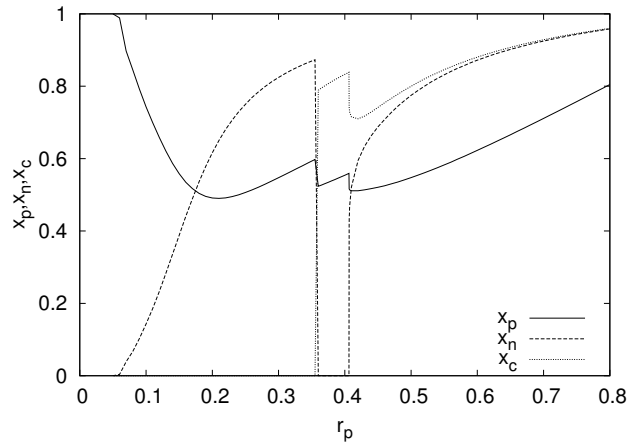}  \vspace{0cm} \caption{ \label{st3rand}Steady-state densities of nutrient ($x_p$), normal ($x_n$), and cancer ($x_c$) cells as  functions of $r_p$ for $d=3$ and $r_c/r_n=1.5$. Simulations were made for the version with random supply of nutrient cells and with $L=50$ and simulation time $t=10^4$.}
\end{figure}

Simulations of such a model for $d=1, 2$ and 3 show that this modification has only a minor effect on the behaviour of the model (Fig.\ref{st3rand}). Indeed, the regions of existence of and coexistence resemble those in the previous $d=3$ version (Fig.~\ref{steadyd3}). We do not present our data but simulations for $d=1$ and 2 also give qualitatively very similar results.

\section{Conclusions}
In the present paper we analysed a simple lattice model of two species competing for common renewable resources.  Although various models have already been examined in the literature, little attention was paid to general properties of such systems as seen from the many-body perspective. The objective of the present paper was to examine a competition of species from this very perspective of statistical mechanics. 
The model might describe the competition  of two predator species for a prey but to some extent also of tumor and normal cells for some nutrients. It is the latter interpretation that was mainly used in our paper.

The introduced  model shows rich behaviour. It exhibits  continuous transitions, which most likely belong to the directed percolation universality class. In some cases, however, the identification of the universality class was inconclusive and it was suggested that a novel critical behaviour might appear in this model. There are also discontinuous transitions in the system. In the $d=1$ version the extinction of tumor cells is always discontinuous, but for $d>1$ versions both continuous and discontinuous extinctions of tumor cells appear. It would be desirable to explain such dimension-dependent behaviour using some more general physical arguments. 

Contrary to the Gause exclusion principle, our model show that in some cases coexistence of predator species that depend on the same prey is possible. In one-dimensional systems coexistence appears only when prey is sufficiently abundant but higher dimension facilitates coexistence. It would be interesting however to generalize such observations examining for example models with interspecific exclusion.

Especially in the context of tumor growth problem it would be desirable to examine in more details the role of the initial state. Preliminary results show that for $d>1$ the final state depends on the initial concentrations of normal and tumor cells, but most likely their spatial distribution plays a role as well. Finding distributions or configurations that can suppress spreading of tumor cells would be particularly valuable.

In the present model, species do not change their behaviour in time. A more comprehensive description of ecosystems as well as tumors \cite{merlo}, in addition to ecological, should take into account also evolutionary aspects.
Analysis of the evolutionary, multi-species extension of the present model might further contribute to a better understanding of such complex, but immensely important systems.
\begin{acknowledgments}
This research was supported with Ministry of Science and Higher Education grant
N~N202~071435. We gratefully acknowledge access to the computing
facilities at Pozna\'n Supercomputing and Networking Center.
\end{acknowledgments}


\begin{thebibliography}{}
\bibitem{comment1} Actually, our model has two absorbing states. One, where all lattice sites are empty and the other where they are occupied by nutrients. However, for the first absorbing state a single nutrient cell will spread and transform it into the second absorbing state. Such a state is thus unstable and dynamically irrelevant. Similar behaviour occurs in some other lattice prey-predator systems \cite{KOWALIK}.
\end{thebibliography}
\end {document}